\documentclass[12pt]{article}
\usepackage[breaklinks=true]{hyperref}
\usepackage{graphicx}
\usepackage{epsfig}
\usepackage{dcolumn}
\usepackage{subfigure}
\usepackage{amsmath,amsthm, amssymb}
\usepackage{comment} 
\usepackage{array}

\newcommand{\bea}{\begin{eqnarray}}
\newcommand{\eea}{\end{eqnarray}}
\newcommand{\beq}{\begin{equation}}
\newcommand{\eeq}{\end{equation}}
\newcommand{\nn}{\nonumber}

\newcommand{\del}{\partial}

\newcommand{\avg}[1]{\left\langle #1 \right\rangle}

\newcommand{\hH}{{\hat{H}}}

\newcommand{\ra}{\rangle}
\newcommand{\la}{\langle}
\newcommand{\vac}{|0\ra}
\newcommand{\cav}{\la 0 |}
\newcommand{\dg}{\dagger}
\renewcommand{\del}{\delta}

\def\be{\begin{equation}}
\def\ee{\end{equation}}
\def\beq{\begin{eqnarray}}
\def\eeq{\end{eqnarray}}

\begin{document}

\begin{center}
\vspace{48pt}
{ \Large \bf On the Quantum Geometry of Multi-critical CDT}

\vspace{40pt}
{\sl Max R. Atkin}$^{a}$
and {\sl Stefan Zohren}$\,^{b,c}$  
\vspace{24pt}

{\small

$^a$~Fakult\"{a}t f\"{u}r Physik, Universit\"{a}t Bielefeld,\\
Postfach 100131, D-33501 Bielefeld, Germany

\vspace{10pt}

$^b$~Department of Physics, Pontifica Universidade Cat\'olica do Rio de Janeiro,\\
Rua Marqu\^es de S\~ao Vicente 225, 22451-900 G\'avea, Rio de Janeiro, Brazil\\
\vspace{10pt}

$^c$~Rudolf Peierls Centre for Theoretical Physics,\\
1 Keble Road, Oxford OX1 3NP, UK

}

\end{center}


\vspace{36pt}

\begin{center}
{\bf Abstract}
\end{center}

We discuss extensions of a recently introduced model of multi-critical CDT to higher multi-critical points. As in the case of pure CDT the continuum limit can be taken on the level of the action and the resulting continuum surface model is again described by a matrix model. The resolvent, a simple observable of the quantum geometry which is accessible from the matrix model is calculated for arbitrary multi-critical points. We go beyond the matrix model by determining the propagator using the peeling procedure which is used to extract the effective quantum Hamiltonian and the fractal dimension in agreement with earlier results by Ambj{\o}rn et al. With this at hand a string field theory formalism for multi-critical CDT is introduced and it is shown that the Dyson-Schwinger equations match the loop equations of the matrix model. We conclude by commenting on how to formally obtain the sum over topologies and a relation to stochastic quantisation.\\ \\
PACS: 04.60.Ds, 04.60.Kz, 04.06.Nc, 04.62.+v.\\
Keywords: quantum gravity, lower dimensional models, lattice models.



\vfill

{\footnotesize
\noindent
$^a$~{email: matkin@physik.uni-bielefeld.de}\\
$^b$~{email: zohren@fis.puc-rio.br}\\
$^c$~{email: zohren@physics.ox.ac.uk}\\
}


\newpage

\section{Introduction}
A long standing problem in the causal dynamical triangulation (CDT) approach to causal quantum gravity (see \cite{Ambjorn:1998xu} for reviews on recent progress in 2D and 4D) has been to obtain an analytical understanding of the behaviour of models that include matter. Shortly after the original proposal of CDT in \cite{Ambjorn:1998xu}, a number of models appeared generalising CDT which exhibited scaling behaviour different from pure CDT \cite{Di Francesco:2001gm}, thereby potentially corresponding to some form of matter coupled to CDT. One short-coming of this early work was that the models did not have a clear flat space analogue and therefore it was difficult to make a comparison to flat space theories or even to matter models coupled to dynamical triangulations (DT), i.e.\ two-dimensional Euclidean quantum gravity. For DT the known analytical results are for the case when the matter sector corresponds to one of the $(p,q)$ minimal conformal field theories (CFTs). It is these CFT models that are of greatest interest, since they may be defined both on non-dynamical lattices and DT. However, until recently models of such CFTs coupled to CDT have resisted an analytical treatment.

Recent progress in \cite{Ambjorn:2012zx,Atkin:2012yt} has changed this situation. In \cite{Ambjorn:2012zx}, the matrix model formulation of CDT proposed in \cite{Ambjorn:2008jf,Ambjorn:2008gk} was generalised to the case of a quartic potential, and it was shown to possess a multi-critical point whose scaling exponents put it in a universality class different from pure CDT. In DT such a matrix model corresponds to hard dimers on a random lattice and therefore to a $(2,5)$ minimal model in the continuum limit. A priori the interpretation of the multi-critical point in the CDT model was less clear; indeed if one adds polygons of arbitrary type to a CDT in a transfer matrix formulation one does not obtain any new critical points. For dimers this can be understood as being due to the fact that each square corresponds to a dimer orientated in two different ways, only one of which respects the casual structure. However, it was argued in \cite{Ambjorn:2012zx} that the new multi-critical point indeed corresponds to dimers coupled to CDT. This argument was confirmed by simultaneous work \cite{Atkin:2012yt} in which a hard dimer model on a CDT was analysed directly via a combinatorial argument based on tree bijections. This givs us confidence that the multi-critical points of the CDT matrix model correspond to the $(2,2p+1)$ CFTs coupled to CDT, as is the case in DT.

Given these developments, some obvious but nonetheless interesting and important quantities characterising the quantum geometry of multi-critical CDT remain to be calculated. In this paper we discuss generalisations of the work in \cite{Ambjorn:2012zx}; a CDT matrix model with an arbitrary potential possess a hierarchy of multi-critical points, the number of which is only limited by the order of the potential. We note that similarly to the case without any matter coupling \cite{Ambjorn:2008jf}, the CDT scaling limit may be taken at the level of the action, resulting in a matrix model formulation of the continuum surface model. At this point we would like to stress that this observation was already made in \cite{Ambjorn:2008gk}. However, in \cite{Ambjorn:2008gk} the resulting expressions were wrongly interpreted as belonging to the same universality class. To clarify this, we redo the analysis and explicitly determine the loop equations and resolvents for an arbitrary multi-critical point. 

Having this relation at hand allows for a whole host of previous results \cite{cap,sft,topo,stochastic} to be easily generalised which is the main content of this paper. In particular, as a generalisation of \cite{cap} we determine the propagator using the peeling procedure in a manner similar to that used in \cite{Watabiki:1993ym} for DT. From this we extract the effective quantum Hamiltonian and the two-point correlation function. Using the latter we compute the Hausdorff dimension $d_H$. We find in agreement with the conjecture in \cite{Ambjorn:2012zx} that $d_H = m/(m-1)$ for the $m$th multi-critical point. Besides the grand-canonical Hasudorff dimension we also calculate the canonical Hausdorff dimension which agrees with the former. Having derived the effective quantum Hamiltonian, we set up a string field theory (SFT) formalism for multi-critical CDT, analogous to the case of CDT without matter coupling \cite{sft} and DT \cite{sft2}. It is shown that the Dyson-Schwinger equations (DSE) of the SFT reproduce the loop equation of the matrix model, which can be used to determine the loop amplitudes order by order in the topological expansion. We conclude by discussing possible extensions related to the sum over topologies and stochastic quantisation in a similar vein to what has been done \cite{topo,stochastic} for CDT without matter.

\section{Multi-critical Points of the CDT Matrix Model}
Consider the one hermitian matrix model,
\beq
\label{CDTmm}
Z = \int [d\phi] e^{-\frac{N}{\beta} \mathrm{Tr} V(\phi) },
\eeq
where $V(\phi) = -g \phi + \frac{1}{2} \phi^2 -\frac{g}{3} \phi^3$. It possesses the loop equation,
\beq
\label{leqn}
\beta W_\phi(z)^2 = V'(z)W_\phi(z) + W^Q_\phi(z),
\eeq
where $W_\phi(z)$ is the resolvent for $\phi$ and $W^Q_\phi(z)$ is a polynomial of order one. Note that since the resolvent corresponds to the disc amplitude with boundary cosmological constant $z$, it can be expressed as $W_\phi(z) = \sum^\infty_{k=0} z^{-k-1} w_l$ where $w_l$ is the disc amplitude for a disc with a boundary of fixed length $l$.

\begin{figure}[t]
\begin{center}
\includegraphics[width=14cm]{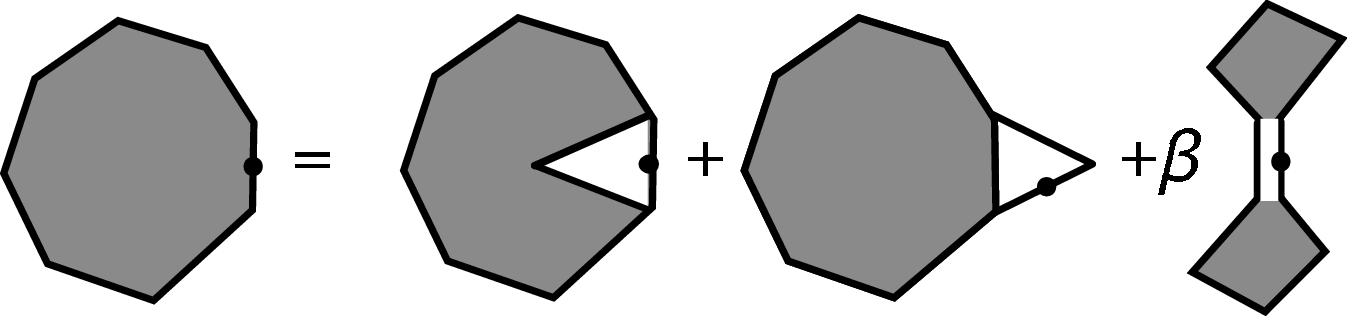}
\caption{Graphical illustration of the loop equation of normal CDT.}
\label{fig1}
\end{center}
\end{figure}

It was first shown in \cite{Ambjorn:2008gk} that \eqref{leqn} can be interpreted as describing the combinatorics of a DT in which spatial topology change i.e. the birth of a baby universe, is assigned a coupling constant $\beta$. Explicitly this is because \eqref{leqn} may be rewritten for large enough $l$ as,
\beq
\label{Lspaceleqn}
w_l = g w_{l+1} + g w_{l-1} + \beta \sum^{l-2}_{m=0} w_{l-m-2} w_m
\eeq
which can be interpreted graphically as shown in Figure \ref{fig1}. Note that the linear term in the potential can be interpreted as the second term on the right hand side of Figure \ref{fig1} \cite{Ambjorn:2008gk}; it adds a triangle (factor of $g$) and increases the boundary by one. Interpreting the loop equation combinatorially we see that if we do not include the linear term the boundary cannot increase once $\beta$ is taken to zero. It is then clear that the term with coefficient $\beta$ is the only one to allow spatial topology change. Hence, by taking $\beta$ to zero the contribution from geometries in which spatial topology change occurs can be suppressed. Furthermore, it was shown that there exists a non-trivial scaling limit about the point $\beta = 0$ corresponding to continuous geometries in which topology change is also assigned a scaled coupling constant $g_s$. It has long been known \cite{Ambjorn:1998xu} that the key difference between DT and CDT is the absence of topology change in the latter and therefore this new scaling limit was interpreted as corresponding to a ``generalised CDT'' in which a controlled amount of topology change is permitted. This interpretation is justified given that known CDT results are recovered upon setting $g_s = 0$.

A particularly interesting feature of the amplitudes obtained for generalised CDT is that they can be written in a form that matches the unscaled functional form of the amplitude, thereby suggesting that the {\emph{continuum quantities can be obtained directly from a matrix model}}. That this is indeed the case was shown in \cite{Ambjorn:2008gk} by demonstrating that the scaling limit associated with CDT could be taken at the level of the action. In particular by substituting into \eqref{CDTmm} the scaling forms,
\beq
\phi = \phi^* \, 1_{N\times N}  + a \Phi, \qquad g = g^* - a^2 \Lambda \qquad \mathrm{and} \qquad \beta = a^3 g_s,
\eeq
where $1_{N\times N}$ is the unity matrix and $\phi^*$, and $g^*$ are defined by requiring $V'(g^*,\phi^*) = V''(g^*,\phi^*) = 0$, we obtain,
\beq
\label{contZ}
\tilde{Z} = \int [d\Phi] e^{-\frac{N}{g_s} \mathrm{Tr} \left[2\Lambda \Phi - \frac{1}{6}\Phi^3 \right] },
\eeq
where $\tilde{Z}$ is the continuum version of $Z$, obtained via a multiplicative renormalisation. The partition function in \eqref{contZ} defined the continuum theory of generalised CDT. Amazingly, since the size of the matrix $N$ in \eqref{contZ} only plays the role of a coupling constant in the perturbative expansion, one can obtain a summation over the entire perturbative string expansion simply by setting $N=1$ and evaluating the above integral \cite{topo}.

In the remainder of this section, we now repeat the analysis of \cite{Ambjorn:2008gk} in detail to generalise this above fact to all multi-critical points, including the first multi-critical point found in \cite{Ambjorn:2012zx,Atkin:2012yt}, where in contrast to \cite{Ambjorn:2008gk} we interpret the higher multi-critical points as new universality classes. One first extends the matrix model so that the potential has the form $V(\phi) = -g \phi + \frac{1}{2} \phi^2 - g \sum^{m+1}_{k=3} \frac{t_k}{k} \phi^k$. This leads to loop equations of the form \eqref{leqn}, where $W^Q_\phi$ is now of order $m-1$, and the corresponding length-space loop equation,
\beq
\label{LspaceleqnGen}
w_l = g w_{l-1} + g \sum^{m+1}_{k=3} t_k w_{l+k-2} + \beta \sum^{l-2}_{m=0} w_{l-m-2} w_m.
\eeq
Note that we let $t_3 = 1$, since in the combinatorial interpretation of the loop equations, shown in Figure \ref{fig1} and generalised above, this corresponds to each triangle having a fugacity of $g$. We therefore have $m-1$ free parameters in the potential.

Although one could solve the loop equation and then look for a scaling limit of the form found for the pure model, it is instead quicker to realise we can again take the scaling limit about $\beta = 0$ at the level of the action. To this end, consider the critical point $(\{t^*_k\}, g^*, \phi^*)$, it is defined by the condition that the derivatives $V^{(n)}(\{t^*_k\}, g^*, \phi^*) = 0$ for $1 \leq n \leq m$. This constitutes $m$ equations for the $m-1$ parameters in the potential together with the quantity $\phi^*$; barring any degeneracy we expect to be able to solve for the critical point. We then substitute the scaling forms,
\beq
\label{scalingforms}
\phi = \phi^* \, 1_{N\times N} + a \Phi, \qquad g = g^* - a^m \Lambda \qquad \mathrm{and} \qquad \beta = a^{m+1} g_s,
\eeq
into the action to obtain,
\beq
\label{contZm}
\tilde{Z} = \int [d\Phi] e^{-\frac{N}{g_s} \mathrm{Tr} \left[(-\partial_g \partial_\phi V^*) \Lambda \Phi + \frac{1}{(m+1)!}(\partial^{m+1}_\phi V^*) \Phi^{m+1} \right] },
\eeq
where we have introduced the notation $V^*$ for $V$ evaluated at the critical point. Again we performed a wave function renormalisation by absorbing multiplicative factors in $\tilde{Z}$. In general, to obtain a non-trivial scaling limit the critical point introduced above should correspond to a point in the parameter space of $Z$ at which $Z$ is non-analytic. For example, in the case of DT the scaling limit corresponds to zooming in on the region near a branch point of $Z$. Here, by taking the scaling limit at the level of the action we have made this point highly implicit, indeed a priori it is not clear the scaling limit defined here will produce a non-trivial result. However, it is clear the integral in \eqref{contZm} resulting from the scaling limit is non-trivial. Furthermore, since we do not scale $N$ in this limit we can see that the scaling limit actually commutes with the matrix integral in \eqref{CDTmm}. To be explicit, \eqref{CDTmm} may be written in terms of the eigenvalues $\lambda_i$ of $\phi$, thus,
\beq
\label{CDTmmeigen}
Z = \int \prod^N_{i =1} d\lambda_i \prod_{j< i} (\lambda_i - \lambda_j)^2 e^{-\frac{N}{\beta} V(\lambda_i) }.
\eeq
The scaling ansatz for $\lambda_i$ is, from \eqref{scalingforms}, $\lambda_i =  \phi^* + a \tilde{\lambda}_i$. Subbing this into \eqref{CDTmmeigen} together with \eqref{scalingforms} we see that the measure is only changed by an overall factor of $a$. We therefore conclude that the small $a$ expansion of \eqref{CDTmmeigen} can be found either by computing the small $a$ expansion of the integrand and then performing the $N$ integrations of the eigenvalues, as we do in this article, or by first computing the integrals over the eigenvalues and then taking the small $a$ expansion of the result as done in \cite{Ambjorn:2008gk,Ambjorn:2012zx}. The advantage of doing the latter is that we see explicitly that this scaling limit corresponds to zooming in on a branchpoint of the partition function (for a disc) while simultaneously taking the branch cut length to zero. Here we are content with commuting the $a \rightarrow 0$ limit through the integral and viewing the resulting non-trivial integrand as evidence that the critical point is non-trivial. Furthermore we see that any scaled amplitudes may be computed from the matrix model \eqref{contZm} without any scaling. 

Let us for simplicity choose the coefficients in the action of \eqref{contZm} such that the potential becomes 
\beq\label{potential}
V(X)= \Lambda X+\frac{1}{m+1}(-X)^{m+1}.
\eeq 
One can now easily derive the loop equation for the resolvent $W(X)$ of the matrix model with the above potential,
\beq\label{loopWX}
g_s W(X)^2 = [\Lambda -(-X)^m] W(X) + W^Q(X),
\eeq
where $W^Q(X)= -(-X)^{m-1} +\sum_{l=0}^{m-2} a_l X^l$ is a polynomial of order $m-1$ and $a_l=a_l(\Lambda,g_s)$. The solution of the loop equation \eqref{loopWX} is formally given by 
\beq\label{loopWXsol1}
W(X) =\frac{-((-X)^m -\Lambda)+\sqrt{((-X)^m -\Lambda)^2 +4 g_s W^Q(X) }}{2g_s}
\eeq
where $W^Q(X)$ is still undetermined. However, one can fix $W^Q(X)$ by noting that \eqref{loopWXsol1} has to obey the one-cut solution
\beq\label{loopWXsol2}
W(X) =\frac{-((-X)^m -\Lambda)+(X^{m-1} +\sum_{l=0}^{m-2} b_l X^l)\sqrt{ (X-c_1)(X-c_2) }}{2g_s},
\eeq
where the $b_l$, $c_1$ and $c_2$ are functions of $\Lambda$ and $g_s$. Matching \eqref{loopWXsol1} and \eqref{loopWXsol2} yields a set of $2m$ equations for the $2m$ unknowns $a_l$, $b_l$, $c_1$ and $c_2$ which completely fix the solution. Solving the set of equations can be tedious for general $m$ and $g_s>0$ and we will not peruse it here. However, we note that in the case of most physical interest of pure multi-critical CDT, i.e.\ when $g_s=0$, the resolvent simplifies to
\beq
W(X) = \frac{W^Q(X)}{(-X)^m-\Lambda}.
\eeq
Demanding that the inverse Laplace transform $W(L)\to0$ as $L\to\infty$ and furthermore that there exist no oscillatory terms in the large $L$ limit,\footnote{This requirement is equivalent to the one-cut assumption.} we find,
\beq\label{WQres}
W^Q(X) =  \sum_{l=0}^{m-1} (-1)^{l+1} \Lambda^{\frac{m-1-l}{m}}  X^l.
\eeq
This yields for the resolvent
\beq
W(X) =\frac{1}{X+\Lambda^{1/m}}.
\eeq
In agreement with the corresponding result for $m=3$ as obtained in \cite{Ambjorn:2012zx}. Furthermore, it can also be shown that the combinatorial approach derived for (restricted) dimers coupled to CDT \cite{Atkin:2012yt} yields the same result with $m=3$.

\section{Finite Time Propagator and Fractal Dimension}
Although the matrix model approach is well adapted to analytically studying different forms of matter coupled to CDT it is not so suited to providing information about the geometry of the spacetime. An important observable encoding more refined properties of the geometry is the finite time propagator $G(x,y;t)$, which is unavailable in the matrix model. The quantity $G(x,y;t)$ corresponds to a summation over manifolds of cylindrical topology in which the entrance and exit loops are separated by a fixed geodesic distance $t$ and with the boundaries carrying the boundary cosmological constants $x$ and $y$ respectively. The finite time propagator is also a necessary ingredient in the construction of the string field theory which is the topic of the next section. In this section we will compute $G(x,y;t)$ and discuss how a naive calculation of the fractal dimension $d_H$ agrees with the value conjectured in \cite{Ambjorn:2012zx} from considerations of multi-critical branched polymers.

To begin our calculation we first introduce $G(l,l';t)$, which is the length-space version of the propagator, in which the boundaries are of fixed length $l$ and $l'$. We then have $G(x,y;t) = \sum_{l=0}^\infty \sum_{l'=0}^\infty x^{-l-1} y^{-l'-1} G(l,l';t)$. The finite time propagator may then be computed, for the case of a general potential in the matrix model, by viewing \eqref{LspaceleqnGen} as a time dependent process, as first suggested in \cite{Watabiki:1993ym}. We first introduce a new boundary into the spacetime, leading to a modification of \eqref{LspaceleqnGen} of the form,
\beq
G(l,l') = g G(l-1,l') + g \sum^{m+1}_{k=3} t_k G(l+k-2,l') + 2 \beta \sum^{l-2}_{m=0} w_{l-m-2} G(m,l'),
\eeq
where $G(l,l')$ is the propagator with no constraint on the distance between the initial and final boundary and the factor of two in front of the topology changing term is due to the ability to add the exit loop to either of the two baby-universes. Now we envision the addition of polygons to this amplitude as a time dependent process, this is known as the peeling procedure. The LHS of the above equation has one additional polygon on its boundary compared to the RHS. If we were to add an entire space-like layer of new polygons to the amplitude, the distance between the initial and final boundary would be increased by one. Hence for large enough $l$ we can approximate the addition of a single polygon by a change in $t$ of $1/l$, leading to,
\bea
\frac{1}{l}\partial_t G(l,l';t) &=& g G(l-1,l';t) -G(l,l';t)\nn\\
&&+ g \sum^{m+1}_{k=3} t_k G(l+k-2,l';t) + 2 \beta \sum^{l-2}_{m=0} w_{l-m-2} G(m,l';t).
\eea
By summing over $l$ and $l'$, 
this can then be written as,
\beq
\partial_t G(x,y;t) = \partial_x \left( (V'(x) -2 \beta W_\phi(x) ) G(x,y;t)\right).
\eeq
Substituting in the scaling relations \eqref{scalingforms} one find that to obtain a non-trivial limit we must have $t$ scaling as $t = a^{m-1} T$. After rescaling $\Lambda$ and $X$ we obtain,
\beq \label{propdif}
\partial_T G(X,Y;T) = -\partial_X \left[ \left((-X)^m -\Lambda + 2 g_s W(X) \right)G(X,Y;T) \right].
\eeq
Note that $G(X,Y;T)$ and the corresponding expression in length space are related by a Laplace transformation
\beq
G(X,Y;T) =\int_0^\infty dL_1 \int_0^\infty dL_2 \,\, e^{-XL_1-YL_2} \, G(L_1,L_2;T).
\eeq
To solve the differential equation \eqref{propdif} one has to set the initial conditions. In particular, one imposes that $G(L_1,L_2;0)$ has only support on $L_1=L_2$. This implies that $G(L_1,L_2;0)$ is given by a sum over $\delta^{(n)}(L_1-L_2)$, $n\geq 0$, where $\delta^{(n)}(\cdot)$ is the $n$-th derivative of the delta function. In the Laplace transform this is equivalent to 
\beq\label{GT0}
G(X,Y;0)=\frac{P(X)}{X+Y},
\eeq
where $P(X)$ is a polynomial in $X$. We will now show that consistency with the results of the previous section requires that $P(X)=\partial_X W^Q(X)$ with $ W^Q(X)$ given in \eqref{WQres}. In particular, we show that \eqref{propdif} together with \eqref{GT0} yields the correct disc function. 

Let us first note that the disc function satisfies $W(X)=\int^\infty_0 G(X,L_2=0;T) \,dT$ with $2g_s\to g_s$, i.e.\ the disc function corresponds to a propagator of arbitrary time with the final boundary shrunken to zero.\footnote{Note that removing the symmetry factor $2g_s\to g_s$ is the same as in pure CDT \cite{sft} and similarly in the DT string field theory \cite{sft2}. It is naturally, keeping in mind that the propagator is not the two-loop function, as is for example explained in \cite{sft}.}
Integrating \eqref{propdif} with respect to $T$ and shrinking $L_2\to 0$, or equivalently $Y\to\infty$, and removing the symmetry factor $2$ one obtains using \eqref{GT0} that
\beq\label{discP}
-\partial_X W^Q(X) = -\partial_X \left[ \left((-X)^m -\Lambda + g_s W(X) \right)W(X) \right]
\eeq
which after integrating with respect to $X$ precisely yields \eqref{loopWX}.

One observes that \eqref{propdif} yields the following Hamiltonian for pure multi-critical CDT, i.e. with $g_s=0$,
\beq
\label{Geqn}
-\partial_T G_0(X,Y;T) = H_0(X) G_0(X,Y;T), \quad H_0(X)= \partial_X  ((-X)^m -\Lambda),
\eeq 
Using the fact that $G(X,Y;T)$ and $G(L_1,L_2;T)$ are related by an inverse Laplace transformation with respect to $X$ and $Y$ one obtains the Hamiltonian in length space
\beq \label{H0L}
H_0(L)= -L \partial_L^{m} +\Lambda L .
\eeq
Note that for $m$ even the Hamiltonian is Hermitian with respect to the measure $d\mu(L)=L^{-1}dL$.

We now turn our attention to the analysis of the fractal structure of the space-time ensemble as characterised by its fractal or Hausdorff dimension. Given that for any value of $g_s$, the production of baby-universes is under control, we expect the fractal structure of such space-times to be equivalent to the $g_s = 0$ case. We therefore focus our attention on the case of pure multi-critical CDT, by setting $g_s = 0$.

Using the method of characteristics to solve the above equation gives,
\beq
\label{Gexplicit}
G_0(X,Y;T) = \frac{\partial_X W^Q(\tilde{X}(T,X)) }{\tilde{X}(T,X)+Y}\frac{(-\tilde{X}(T,X))^m-\Lambda}{(-X)^m-\Lambda} 
\eeq
with
\beq
\frac{d \tilde{X}(T,X)}{dT} = -((-{\tilde{X}(T,X)})^m - \Lambda),
\eeq
and the condition $\tilde{X}(0) = X$. One can now shrink the initial and final boundary to zero by taking $X$ and $Y$ to infinity, resulting in the amplitude $G_0(T)$ for a sphere with two punctures separated by a distance $T$,
\beq
G_0(T) =\partial_X W^Q(\tilde{X}(T))   \left[(-\tilde{X}(T))^m-\Lambda\right],
\eeq
where we set $\tilde{X}(T)=\tilde{X}(T,X=\infty)$. One now has 
\beq\label{eqT}
T=\int^{\infty}_{\tilde{X}(T)} \frac{dX}{(-X)^m - \Lambda}.
\eeq
From this expression one already sees that one has the following scaling
\beq
G_0(T) = \Lambda^{2 \frac{m-1}{m}} F_1(T \Lambda^{\frac{m-1}{m}}),
\eeq
in terms of a yet underdetermined function $F_1$.
This scaling shows that area scales as $A\sim T^{m/(m-1)}$ and thus that the Hausdorff dimension is 
\beq
d_H=\frac{m}{m-1}
\eeq
in agreement with the value of \cite{Ambjorn:2012zx} suggested from the comparison with multi-critical branched polymers \cite{multiBP}, as well with a recently introduced model of multi-critical tensor models \cite{Bonzom:2012sz}. This dimension can be seen as a ``grand-canonical'' definition of the Hausdorff dimension \cite{scaling}. One can also define the ``canonical'' Hausdorff dimension $d_h$. To do so consider the volume of a spherical shell $\avg{L(T)}_V$ at fixed volume $V\gg T^{d_H}$. The ``canonical'' Hausdorff dimension is then defined through $\avg{L(T)}_V\sim T^{d_h-1}$. In particular, one expects the following scaling \cite{scaling,cm2}
\beq\label{scalingrelation}
\avg{L(T)}_V= V^{1-1/d_H} F_2 (T/V^{1/d_H}),
\eeq
where $F_2(x)\sim x^{d_h-1}$ as $x\to0$ and $F_2(x)\sim \exp(-x^{d_H/(d_H-1)})$ as $x\to\infty$. For pure CDT one has for example $d_H=d_h=2$ and for DT $d_H=d_h=4$. However, for multi-critical branched polymers it is known that the both dimensions differ and one has $d_H=m/(m-1)$, while $d_h=2$ \cite{multiBP,cm2}. We have already seen that for multi-critical CDT we have $d_H=m/(m-1)$ in agreement with the multi-critical branched polymers. In this comparison it is therefore interesting to see whether for multi-critical CDT one has the same behaviour.
We can obtain the Laplace transform of  $\avg{L(T)}_V$ from the propagator \eqref{Gexplicit}. In particular, taking the first boundary to zero ($X\to\infty$), one gets
\beq
\avg{L(T)}_\Lambda = -\left[ \frac{1}{G(Y;T)} \frac{\partial }{\partial Y} G(Y;T)\right]_{Y=0} = \tilde{X}(T)^{-1}.
\eeq
For small $T\ll 1/\Lambda^{1/d_H}$ one has the following expansion from the characteristic equation \eqref{eqT},
\beq
\avg{L(T)}_\Lambda \sim T^{\frac{1}{m-1}} -c_m \Lambda T^{\frac{m+1}{m-1}}+ ...
\eeq
However, from the Laplace transform of \eqref{scalingrelation} one expects that $\avg{L(T)}_\Lambda \sim T^{d_h-1}$ for $T\ll 1/\Lambda^{1/d_H}$ and thus we conclude that $d_h=d_H=\frac{m}{m-1}$. This means that the continuum model has the same fractal properties at all scales. This is probably what one would expect for a continuum surface model which are not simply trees as in the case of multi-critical branched polymers. 

\section{String Field Theory}

In this section we introduce a string field theory (SFT) formalism for multi-critical CDT analogous to the SFT for generalised CDT \cite{sft}, i.e.\ $m=2$. A previous construction for DT can be found in \cite{sft2}.

Let us start by defining a vacuum $\vac$ and creation and annihilation operators with the following properties
\beq\label{s1} 
[\Psi(L),\Psi^\dg(L')]=L\delta(L-L'),~~~\Psi(L)\vac = \cav \Psi^\dg(L) =0. 
\eeq
One then defines the ``second-quantised'' Hamiltonian
\beq
\hH_0 = \int_0^\infty \frac{dL}{L} \; \Psi^\dg (L) H_0(L) \Psi(L),
\eeq
where $H_0(L)$ is the first-quantised Hamiltonian of pure multi-critical CDT as given in \eqref{H0L}.
From this one can rewrite the propagator $G (L_1,L_2;T)$ for pure multi-critical CDT as
\beq
G_0 (L_1,L_2;T) = \cav \Psi(L_2) e^{-T \hH_0} \Psi^\dg(L_1) \vac,
\eeq 
i.e.\ as the creation of a string of length $L_1$, a propagation over time $T$ with respect to $H_0(L)$ and the annihilation of the string given that it has length $L_2$.

\begin{figure}[t]
\begin{center}
\includegraphics[width=14cm]{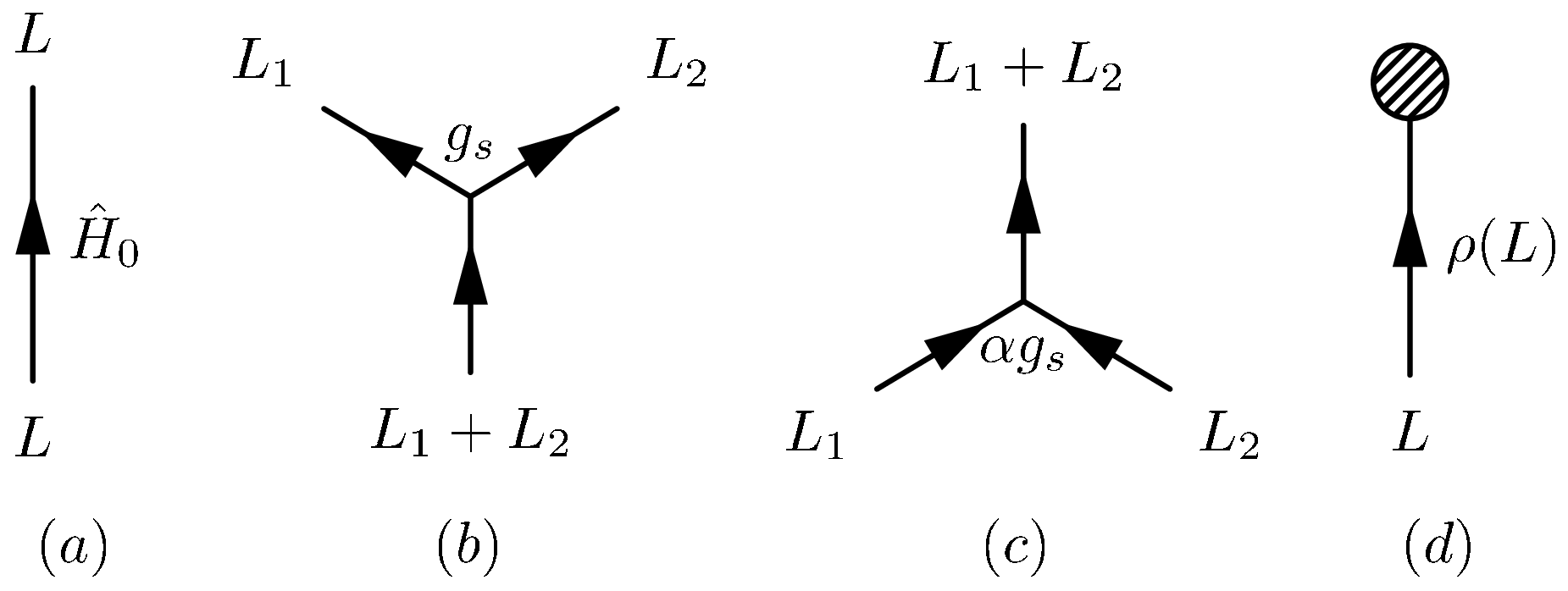}
\caption{The Feynman rules of the string field theory: (a) represents the propagation of a single string with no spatial topology change, (b) and (c) correspond to the splitting and merging of strings respectively and (d) is a tadpole diagram representing the process of a string disappearing into the vacuum.}
\label{fig2}
\end{center}
\end{figure}

We can now set up the SFT by allowing strings to split, merge and be annihilated. This is implemented using the following Hamiltonian
\bea
\hat{H} = \hat{H}_0\!\!\!\!\! \!\!\!&&-~ g_s \int dL_1 \int dL_2 \Psi^\dg(L_1)\Psi^\dg(L_2)\Psi(L_1+L_2) \nn\\
 && -\, \alpha \,g_s\int dL_1 \int dL_2 \Psi^\dg(L_1+L_2)\Psi(L_2)\Psi(L_1)
-\int \frac{dL}{L} \; \rho(L) \Psi(L), \label{HSFT}
\eea
as illustrated in Figure \ref{fig2}. We introduced the coupling $g_s$ for each splitting and joining of the string as well as a factor of $\alpha$ controlling the joining process and thus the topology of the string world-sheet. In relation to the matrix model formulation it will be clear that $\alpha=1/N^2$, i.e.\ $\alpha=0$ corresponds to the planar or large-$N$ limit. Finally, $\rho(L)$ is the tadpole term related to the annihilation of a string. One expects $\rho(L)$ to only have support on $L=0$ and furthermore it is natural to have 
\beq
\rho(L)=G(L,0;0),
\eeq
where $G(L,L_2;T)$ is the propagator as derived in the previous section.

Given the STF Hamiltonian \eqref{HSFT}, one can now use the field theoretical formalism (see \cite{sft,sft2}) to derive $n$-loop amplitudes by introducing a source term in the Hamiltonian yielding
\beq\label{disconn}
\lim_{T\to\infty} \cav e^{-T \hH} \; \Psi^\dg(L_1)\cdots \Psi^\dg(L_n)\vac = 
\left.\frac{\delta^n Z(J)}{\delta J(L_1)\cdots \delta J(L_n)}\right|_{J=0},
\eeq
where the generating function $Z(J)$ is given by
\beq\label{ZJ}
Z(J)= \lim_{T\to\infty}\cav e^{-T \hH} \; e^{\int dL \, J(L) \Psi^\dg(L)}\vac.
\eeq
The amplitudes $\lim_{T\to\infty} \cav e^{-T \hH} \; \Psi^\dg(L_1)\cdots \Psi^\dg(L_n)\vac$ still includes disconnected world-sheets. The amplitudes for the connected world-sheets are obtained from the generating function
\beq
F(J)=\log Z(J)
\eeq
in a similar manner
\bea
W(L_1,...,L_n)&=&\alpha^{1-n}\lim_{T\to\infty} \cav e^{-T \hH} \; \Psi^\dg(L_1)\cdots \Psi^\dg(L_n)\vac_{con.} \nn\\
&=& \alpha^{1-n}
\left.\frac{\delta^n F(J)}{\delta J(L_1)\cdots \delta J(L_n)}\right|_{J=0}.\label{connected}
\eea
Here $W(L_1,...,L_n)$ is the so-called $n$-loop amplitude. Note that we introduced the coupling $\alpha$ in the SFT Hamiltonian \eqref{HSFT} precisely to keep track of the topology of the string world-sheet. However, a genus zero $n$-loop amplitude naturally comes with $n-1$ mergers of the string. To compensate for this we introduced the factor of $\alpha^{1-n}$ in the definition of $W(L_1,...,L_n)$.

It was shown in \cite{sft} that commuting the $\Psi(L)$'s in \eqref{ZJ} past the source term yields the following equation for $F(J)$,
\bea
0= \int_0^\infty dL \, J(L)
\left\{  H_0(L)\, \frac{\del F(J)}{\del J(L)} - \rho(L) 
 -g_s L \int_0^L dL'\;\frac{\del^2 F(J)}{\del J(L')\del J(L-L')} \right. 
\nonumber\\
\left. -g_s L \int_0^L dL'
\frac{\del F(J)}{\del J(L')}\frac{\del F(J)}{\del J(L-L')}
-\alpha g_s L\int_0^\infty dL' L'J(L')\frac{\del F(J)}{\del J(L+L')}\right\}.
\eea
Taking derivatives with respect to $\delta/\delta J(L)$ of this equation yields the Dyson-Schwinger equations (DSE) of the SFT. In particular, taking the first derivative with respect to $J(L)$, setting $J(L)=0$ and Laplace transforming we obtain the DSE for the disc function,
\beq\label{DSE}
0= H_0(X) W(X) -\partial_X W^Q(X) +g_s \partial_X( \alpha W(X,X) + W(X)W(X)).
\eeq
Setting $\alpha=0$, i.e.\ in the planar limit, this equation precisely yields the loop equation \eqref{loopWX} derived from the matrix model with potential \eqref{potential} in the planar limit. Furthermore, including the $1/N^2$ contribution in the loop equation, \eqref{loopWX} becomes
\beq \label{loopWXN2}
g_s ( W(X)^2 + \frac{1}{N^2} W(X,X))= V'(X)W(X) + W^Q(X).
\eeq
Hence, differentiating the loop equation \eqref{loopWXN2} with respect to $X$ one obtains precise agreement with the DSE equation \eqref{DSE} if we identify $\alpha=1/N^2$, as was also the case for the SFT of CDT without matter coupling \cite{sft}. Furthermore, one can convince oneself that the same equivalence between the DSE of the SFT on the one hand and the loop equations of the matrix model on the other hand still holds true when considering higher loop amplitudes. 

\section{Discussion}
In this paper we discussed how the multi-critical CDT scaling limit of the one matrix model found in \cite{Ambjorn:2012zx} can be generalised to obtain a hierarchy of multi-critical points completely analogous to the multi-critical points of the DT scaling limit as obtained in \cite{Ambjorn:2008jf,Ambjorn:2008gk}. In the DT case these multi-critical points correspond to a continuum gravity model coupled to minimal CFT matter of $(2,2p+1)$ type. 
Based on the analysis in \cite{Ambjorn:2012zx,Atkin:2012yt} it is highly likely that a similar interpretation can be given to the higher order critical points.

In going beyond the matrix model, we have obtained via a peeling produce the differential equation satisfied by the finite time propagator and hence the effective quantum Hamiltonian for the time evolution of the space-like universe. This allowed us to generalise many of the results for generalised CDT without matter, already present in the literature. Firstly we have investigated the conjecture in \cite{Ambjorn:2012zx} that the Hausdorff dimension (fractal dimension) of the multi-critical CDT is $d_H = m/(m-1)$ both in the grand-canonical as well as canonical sense. Secondly, we have shown that a string field theory can be built using this Hamiltonian, which in principle allows for the calculation of any amplitude with a specified number of baby-universes and handles. The structure of this SFT is similar to the SFT constructed to describe the multi-critical non-critical string. However, a pleasing aspect of the construction here is that the tadpole term can be naturally identified with the finite time propagator between a universe of finite size and of zero size over an interval of zero time. This is in contrast with the DT case in which the tadpole term had to be selected by hand.

An important quality of the multi-critical CDT scaling limit is that the limit may be taken at the level of the matrix integral. That this can be done is particularly interesting, as the Dyson-Schwinger equations can be identified with the loop equations of the ``continuum'' matrix model provided we identify the coupling $\alpha$ appearing in the SFT for the merging of two universes with $1/N^2$, where $N$ is the size of the matrix in the scaled matrix model. This is an extremely important point as it means the Feynman diagrams appearing in the large $N$ expansion of the continuum matrix model have a direct interpretation as processes in the SFT. The only role of $N$ in the continuum matrix model is to modify the coupling associated to the merging of two universes, thereby providing an independent weight associated to adding a handle to the worldsheet. If we do not care to distinguish between the processes of merging and splitting of universes then we are free to set $N=1$ in the the continuum matrix integral. This is particularly profound as it reduces the matrix integral to a simple integral of one variable. We therefore have that the entire non-perturbative expression for the continuum partition function is,
\beq
\label{nonpertZ}
\tilde{Z}(\Lambda) = \int dz e^{- \frac{1}{g_s} (\Lambda z + \frac{1}{m+1} (-z)^{m+1} )}.
\eeq
That this is indeed the case can likely be obtained by considering CDT in the framework of stochastic quantisation, which we expect to yield the generalised Hamiltonian,
\beq
H(L)= -L \partial_L^{m} +\Lambda L-g_s L^2 . 
\eeq
This would generalise the Hamiltonian, which includes topology change, found for the case of pure CDT using this approach in \cite{stochastic}. The solutions to the Wheeler-de-Witt equation in that case do indeed correspond to the expression obtained by evaluating \eqref{nonpertZ} for the case $m=2$. This is something we hope to pursue in future work.

Finally, it is interesting to note the connections between the work here and the results of \cite{Hashimoto:2005bf}. In \cite{Hashimoto:2005bf} the continuum limit of the exact partition function for a stack of $N$ FZZT branes in the $(1,m)$ background was computed and was found to be given by,
\beq
Z = \int [d\Phi] e^{-\mathrm{Tr}\left[ \Lambda\Phi - \frac{1}{m+1}\Phi^{m+1}  \right] },
\eeq
where $\Phi$ is an $N \times N$ hermitian matrix and $\Lambda$ is a diagonal matrix whose entries are the boundary cosmological constants of each brane. The continuum limit of $m$th order multi-critical CDT is therefore mathematically equivalent to the theory of a stack of $N$ FZZT branes in a $(1,m)$ background with each brane carrying an identical boundary cosmological constant. This also relates to a similar observation made in \cite{Ambjorn:2009rv} for the case of pure CDT. The physical origin of this picture is certainly something that should be better understood.

\subsection*{Acknowledgements} The authors would like to thank J. Ambj{\o}rn, G. Giasemidis, B. Niedner and J. Wheater for discussions, as well as the anonymous referee for comments improving the manuscript. MA acknowledges the financial support of Universit\"{a}t Bielefeld. SZ acknowledges financial support of the STFC under grant ST/G000492/1. Furthermore, he would like to thank the Mathematical Physics Group at Universit\"{a}t Bielefeld for kind hospitality and financial support for a visit during which this work was initiated.
 


\providecommand{\href}[2]{#2}\begingroup\raggedright\endgroup

\end{document}